\documentclass[twocolumn,showpacs,preprintnumbers,amsmath,
 prb,amssymb]{revtex4}
\usepackage{graphicx}
\usepackage{dcolumn}
\usepackage{bm}
\usepackage{longtable}
\begin{document}
\title {Magnetism in C$_{60}$ Films Induced by Proton Irradiation}
\author{S. Mathew, B. Satpati, B. Joseph and B. N. Dev}
\email{bhupen@iopb.res.in}
\affiliation{Institute of Physics, Sachivalaya Marg, Bhubaneswar-751 005, India}
\author{R. Nirmala and S. K. Malik}
\affiliation{Tata Institute of Fundamental Research, Mumbai 400 005, India}
\date{\today}

\begin{abstract}
It is shown that polycrystalline fullerene thin films on hydrogen
passivated Si(111) substrates irradiated by 2 MeV protons displays
ferromagnetic-like behavior at 5 K. At 300 K both the pristine and
the irradiated film show diamagnetic behavior. Magnetization data
in the temperature range 2 - 300 K, in 1 Tesla applied field, for
the irradiated film shows much stronger temperature dependence
compared to the pristine film. Possible origin of
ferromagnetic-like signals in the irradiated films are discussed.
\end{abstract}

 \maketitle

Recent observation of occurrence of ferromagnetism in materials
purely of carbon origin \cite{Esqui_arixv} has created enormous
interest in these materials. Since 1991, it is known that some
organic molecules with unpaired $\pi$ electrons show magnetic
order below 20 K \cite{pi-ele-mag}. An antiferromagnetic ordering
has been found in a $s-p$ electron system below 50 K
\cite{prl_80}. Makarova {\it et al.} \cite{Mak_nat} and Wood {\it
et al.} \cite{Wood_jpcm} showed the occurrence of ferromagnetic
ordering in two dimensionally polymerized highly oriented
rhombohydral C$_{60}$ (Rh-C$_{60}$) phase.
Tetrakis(diemethylamino)ethylene (TDAE) doped fullerides are known
to be ferromagnetic below a Curie temperature (T$_{c}$) of 17 K
\cite{tdae-1,tdae-2}. Ferromagnetic behavior has also been
observed in micro crystalline carbon \cite{pryo_C} and micro
graphite structures \cite{micro_C}. Theoretical efforts in
understanding magnetism in C-based systems are being made. A
ferromagnetic phase of mixed $sp^2$ and $sp^3$ pure carbon has
been predicted theoretically \cite{Fe_mag_theory}. Understanding
the basic mechanism behind magnetic behavior of carbon-based
materials and engineering novel ferromagnetic carbon structures
are of prime importance.

Recently, Esquinazi {et al.} \cite{Esqu_prl91} reported
ferromagnetic (or ferrimagnetic) ordering of highly oriented
pyrolitic graphite (HOPG) with T$_c$ above 300 K when irradiated
with 2.25 MeV protons. The possibility of occurrence of
ferromagnetic, antiferromagentic and superconducting instabilities
due to topological disorder in graphine sheets has also been
predicted theoretically \cite{topology}. It has been suggested
that topological defects can be used to explain the observed
ferromagnetism in Rh-C$_{60}$. According to a recent theoretical
investigation by Vozmediano {\it et al.} \cite{theory_arxiv},
proton irradiation can produce large local defects which give rise
to the appearance of local moments whose interaction can induce
ferromagnetism in a large portion of the graphite sample.

The discovery of ferromagnetism in rhombohedral C$_{60}$ polymers
has opened up the possibility of a whole new family of magnetic
fullerenes and fullerides. Here we present the results of
magnetization measurements on pristine and 2 MeV proton irradiated
C$_{60}$ films deposited on hydrogen passivated Si(111) surfaces.
We find that magnetism is induced in proton irradiated C$_{60}$
films. Transmission electron microscopy (TEM), micro Raman,
nuclear resonant scattering (NRS) and proton induced X-ray
emission (PIXE) measurements were used to characterize the
samples. Magnetic measurements were carried out using a
superconducting quantum interference device (SQUID) magnetometer.

The hydrogen passivation of Si(111) surfaces involves the
following: degreasing of the Si substrate, removal of native
oxide, and then growth of a thin uniform oxide, etching the oxide
produced and passivating the surface with H, using 40\% NH$_4$F
solution. The detailed procedure is given in ref. \cite{H-pass}.
The passivated substrates [H-Si(111)] were loaded into a high
vacuum (4 $\times$ 10$^{-6}$ mbar) chamber and deposition of the
fullerene film was carried out by evaporating 99.9\% pure C$_{60}$
[MER Inc., USA] from a tantalum boat.

Nuclear resonant scattering (NRS) measurements and ion irradiation
were carried out using the 3 MV 9SDH2 tandem Pelletron accelerator
facility in our laboratory. We usually use Rutherford
backscattering spectrometry (RBS) experiments to determine the
film thickness. However for C the Rutherford scattering cross
section is rather small. To enhance the scattering cross section,
we choose appropriate energy of the incident ions and use a
resonant scattering condition. A beam of 4.265 MeV alpha particles
was used for the NRS experiments to determine the film thickness.
These measurements were carried out on a sample area of 8 mm$^2$
and thickness is estimated to be 1.88 $\mu$m from NRS. Then the
C$_{60}$ films were irradiated uniformly with a 2 MeV H$^+$ beam
by rastering the ion beam on the sample. The ion fluence of the
irradiated sample was 6 $\times$ 10$^{15}$ H$^{+}$/cm$^{2}$. The
beam current during irradiation was kept at $\sim$ 50 nA. Plan
view TEM measurements were carried out using 200 keV (JEOL 2010)
HRTEM with point to point resolution of 0.19 nm and lattice
resolution of 0.14 nm. Raman spectra were recorded at room
temperature using a Joven Yvon Raman spectrometer with Ar laser
(514 cm$^{-1}$, 25 mW-cm$^{-2}$). Magnetization was measured using
the reciprocating sample option (RSO) in a SQUID magnetometer
(MPMS XL, Quantum Design) in the temperature range of 2 - 300 K in
applied fields up to 7 Tesla.

\begin{figure}[htbp]
\begin{center}
\includegraphics[width=1.15\columnwidth]{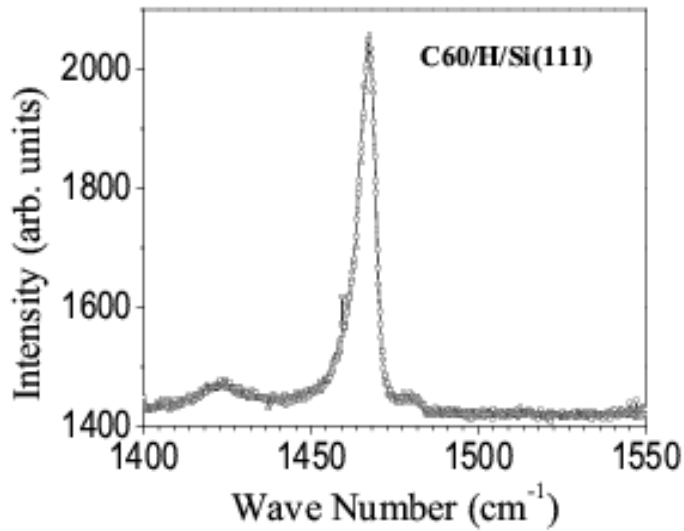}
\caption{A Raman spectrum of C$_{60}$/H-Si(111) film taken at  room
temperature.}
\label{Raman}
\end{center}
\end{figure}

A Raman spectrum from an as-deposited C$_{60}$ film [Fig. 1] shows
a strong peak at 1467 cm$^{-1}$ which corresponds to the
characteristic A$_{g}$ mode of C$_{60}$ \cite{sahoo}. High
resolution TEM (HRTEM) lattice images of as-deposited and
irradiated samples with corresponding diffraction patterns are
shown in Fig. 2. The transmission electron diffraction(TED) and
HRTEM images confirm the crystalline nature of the C$_{60}$ film
in the fcc structure. Twin structures are seen in Fig. 2(a).
Presence of defect structures in the irradiated film is evident
from Fig 2(c). The (111) planar spacing of an fcc fullerene film
is seen in Fig 2 (a) and (c).

\begin{figure}
\includegraphics[width=2.8in]{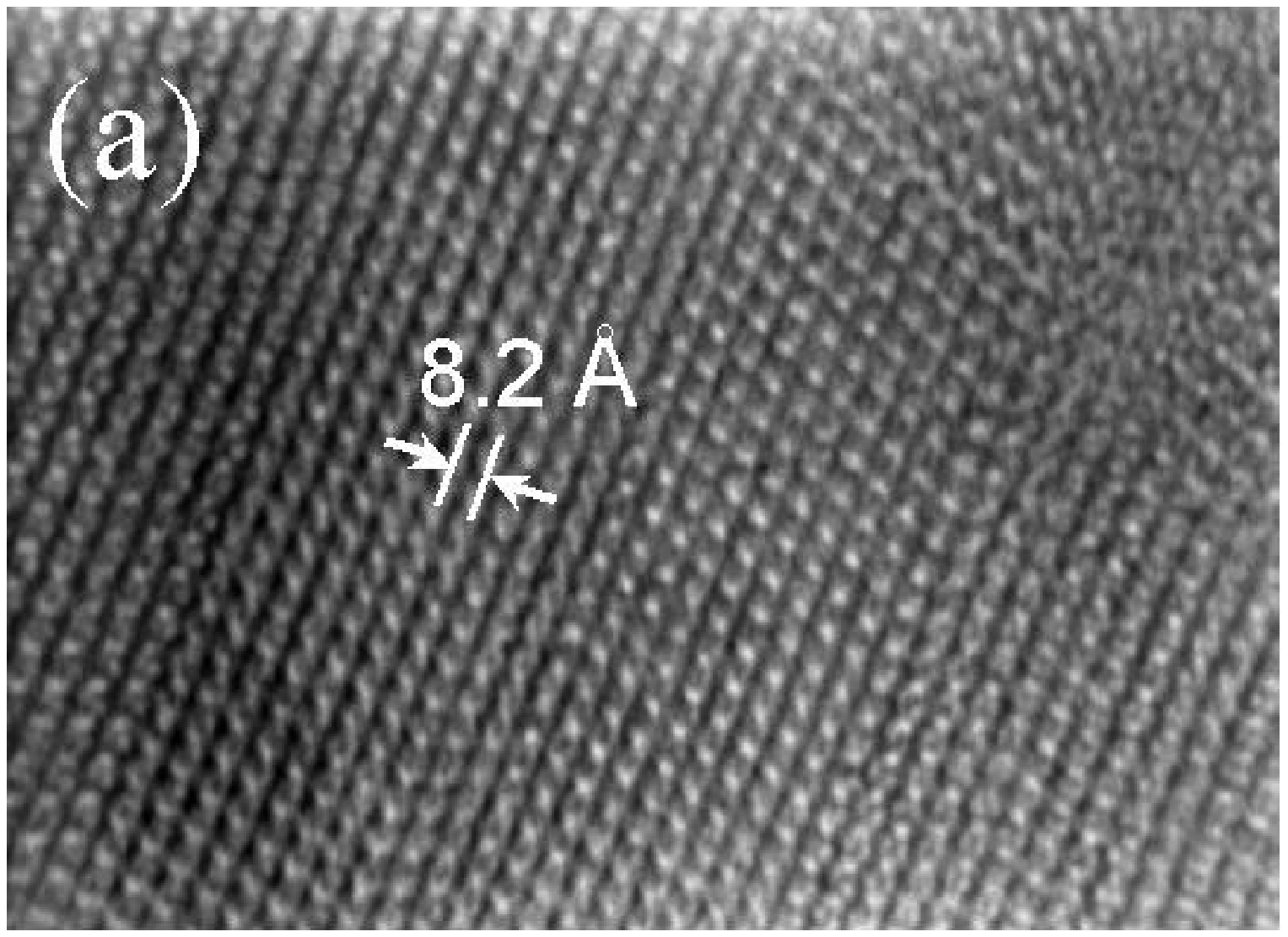}\\
\vskip 0.25cm
\includegraphics[width=2.8in]{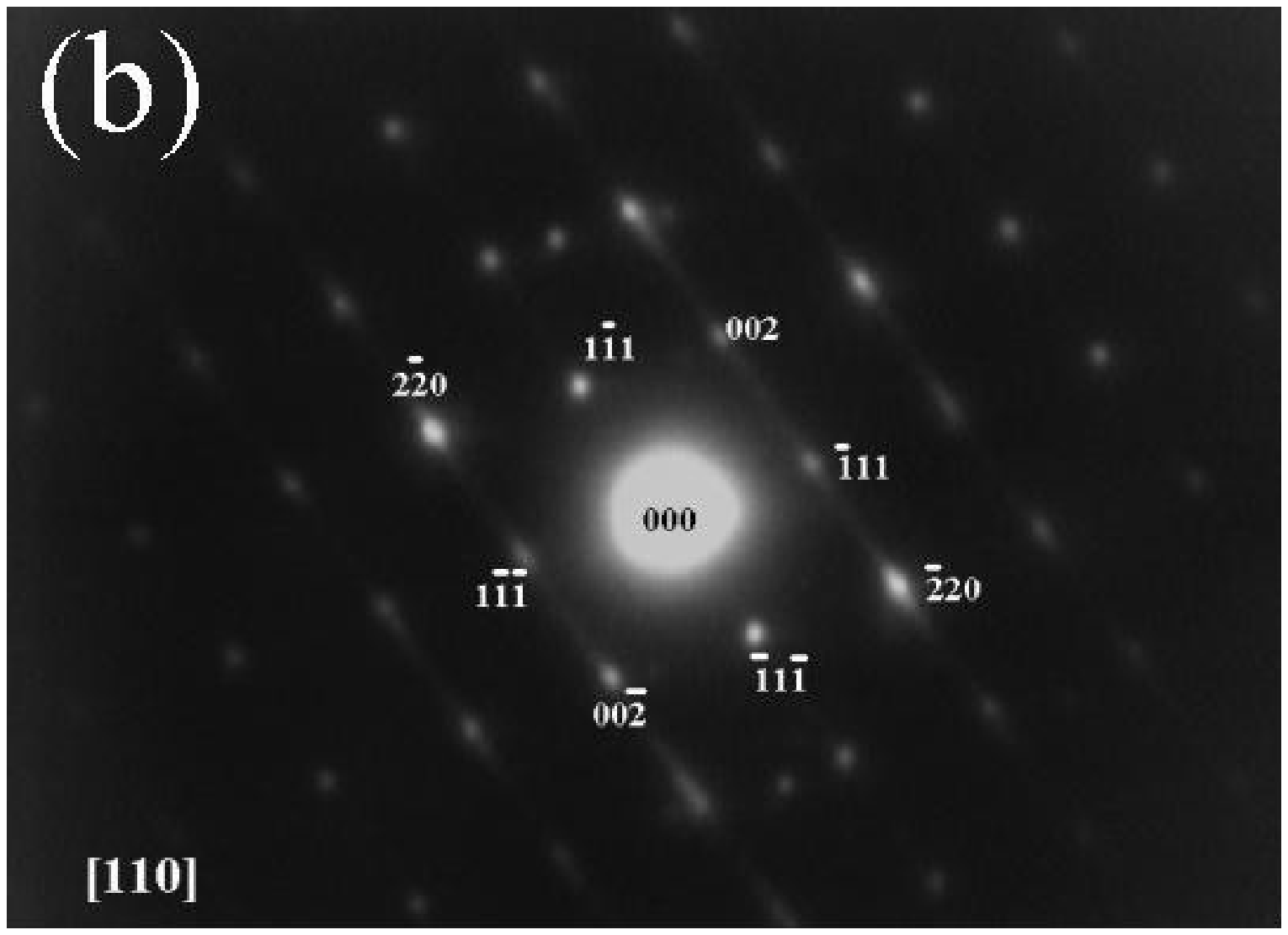}\\
\vskip 0.25cm
\includegraphics[width=2.8in]{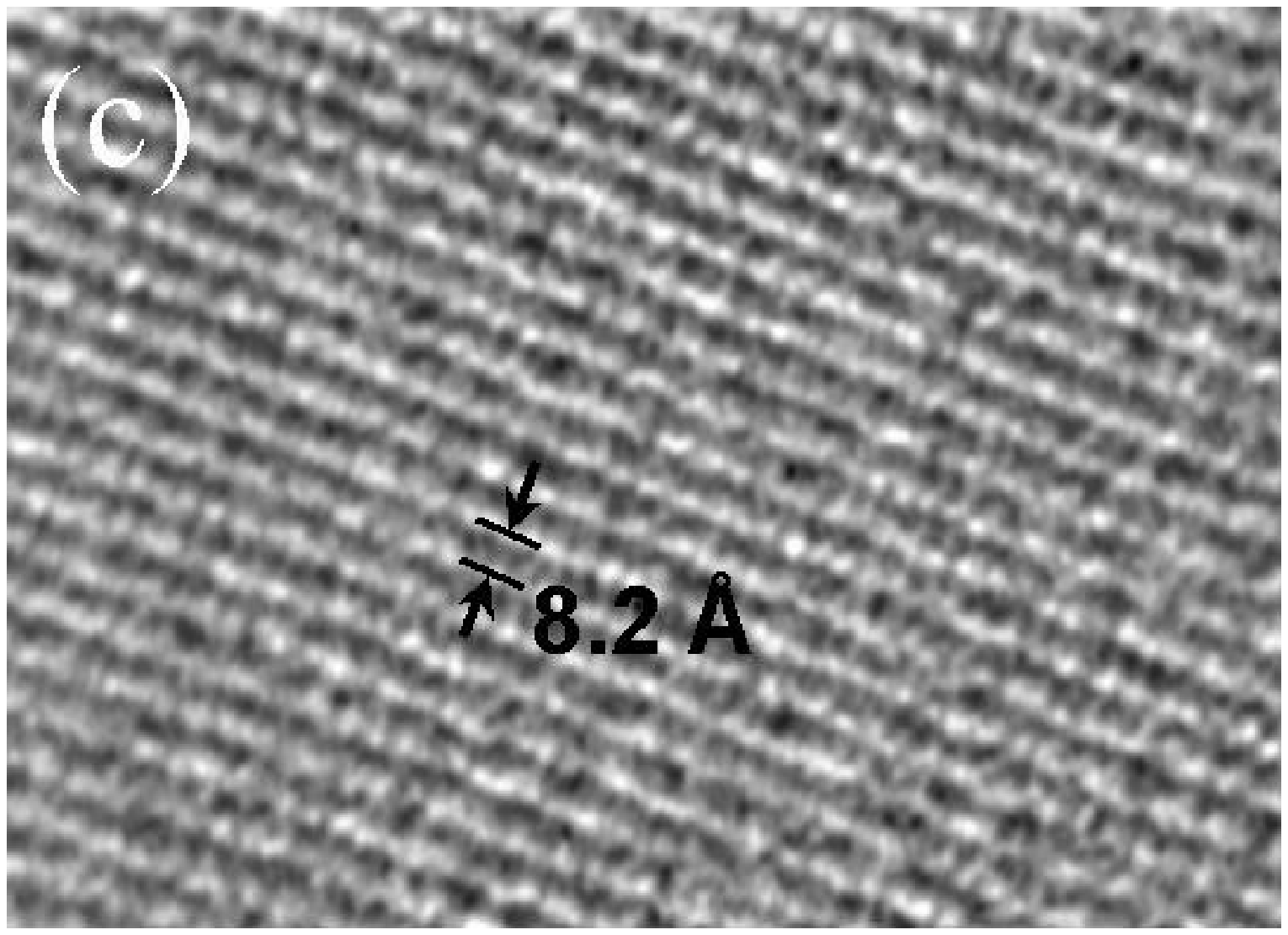}\\
\vskip 0.25cm
\includegraphics[width=2.8in]{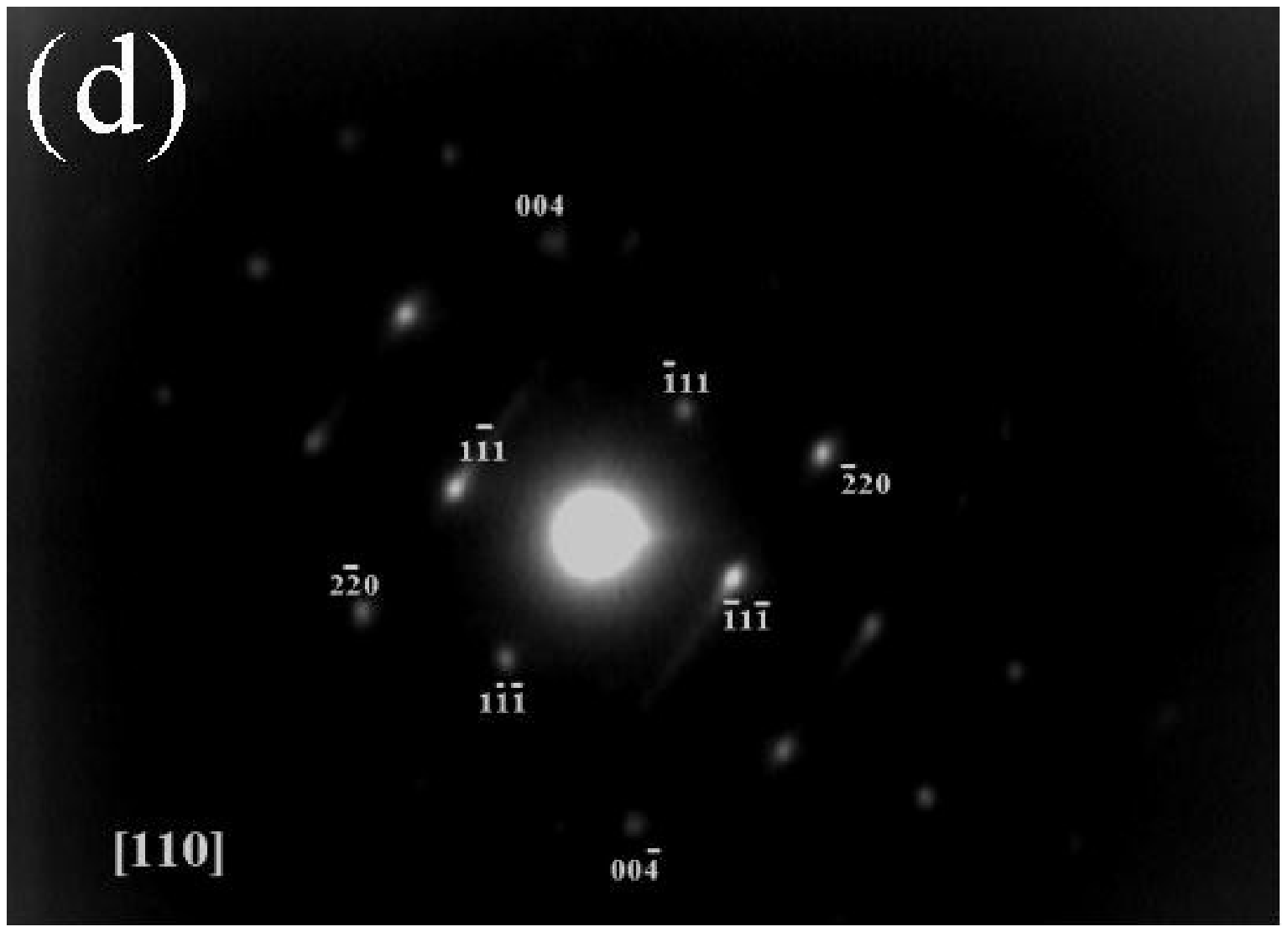}
\caption{ (a) A high resolution TEM (HRTEM) lattice image from an
as-deposited C$_{60}$ film and (b) the corresponding selected area
transmission electron diffraction (TED) pattern; (c) A HRTEM image
from an irradiated C$_{60}$ film and (d) the corresponding TED
pattern. The (111) planar spacing of fcc fullerene is marked in
(a) and (c).} \label{agpb-ted}
\end{figure}

Results of magnetization vs. field (M-H) measurements at 5 K for
the film irradiated at a fluence of 6 $\times$ 10$^{15}$
H$^+$/cm$^2$ show a marked increase of magnetization and a
tendency towards saturation [Fig. 3]. A weak remanent
magnetization of the order of a few tens of $\mu$emu is observed
(data not shown). At 300 K, both the as-deposited and the
irradiated sample show diamagnetic behavior. Magnetization data in
the temperature range of 2 - 300 K, in 1 T applied field, for the
irradiated film shows much stronger temperature dependence when
compared with that of the pristine film [Fig. 4].

\begin{figure}[htbp]
\begin{center}
\includegraphics[width=0.95\columnwidth]{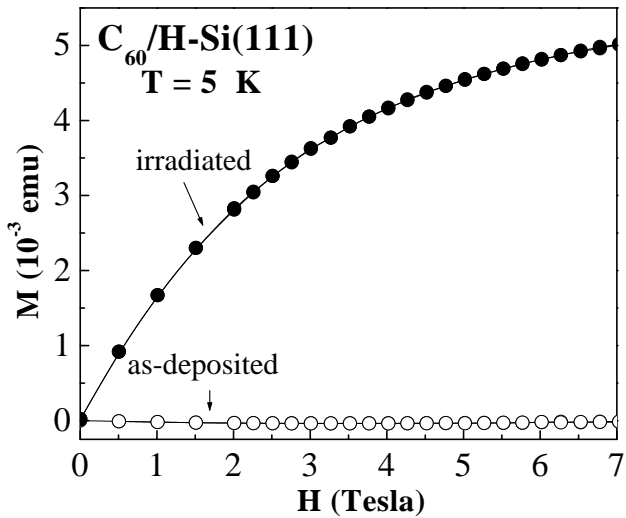}
\caption{M vs. H at 5 K for an as-deposited and an irradiated
C$_{60}$ film after subtracting the substrate [H-Si(111)]
contribution (substrate data not shown here). } \label{magneti}
\end{center}
\end{figure}

The magnetic moment observed for the irradiated sample in an
applied field of 7 T is about 5$\times$ 10$^{-3}$ emu with a
tendency towards saturation. Magnetization curve of the irradiated
film in high fields at 5 K compared with that of the as-deposited
film  and the empty substrate (the latter not shown here) gives
clear evidence for the irradiation-induced magnetism in C$_{60}$
films. The total amount of magnetic impurities (Fe, Cr, Ni) was
determined by post-irradiation proton induced X-ray emission
(PIXE) experiments and estimated to be $\sim$ 50 ppm and the
maximum magnetic moment contribution due to all these impurities
in our film will be less than 5 $\times$ 10 $^{-7}$ emu. Thus the
contribution to observed magnetization due to these impurities is
negligible. Saturation magnetization has been observed in pressure
polymerised C$_{60}$ films in applied fields of less than 2 T
\cite{Mak_nat} whereas in the present system complete saturation
is not attained even in an applied field of 7 T.

\begin{figure}[htbp]
\begin{center}
\includegraphics[width=0.95\columnwidth]{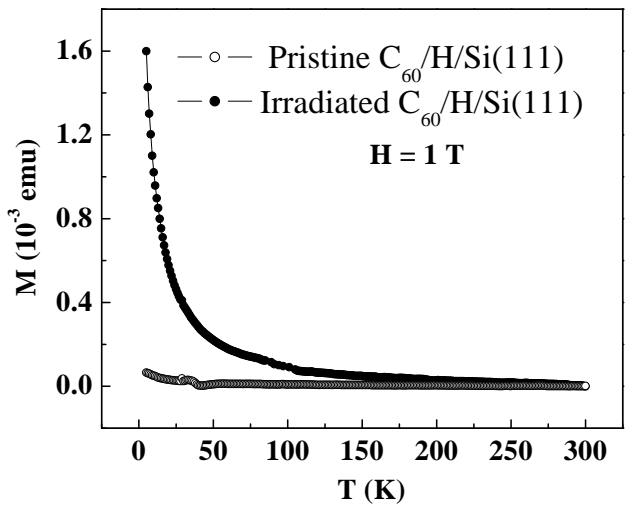}
\caption {M vs. T for an as-deposited and an irradiated C$_{60}$
film in an applied field of 1 T (substrate [H-Si(111)]
contribution subtracted).} \label{MT}
\end{center}
\end{figure}

The range of 2 MeV protons, calculated by using SRIM \cite{srim}
simulation programme for an amorphous carbon target having the
density of C$_{60}$, is found to be $\sim$ 50 $\mu$m. Since our
C$_{60}$ film is only $\sim$ 1.9 $\mu$m thick , the protons pass
through the film and get buried deep into the Si substrate. The
total energy loss of the proton beam in the present 1.9 $\mu$m
thick C$_{60}$ film is $\sim$ 45 keV.  The energy loss of protons
at the top and the bottom of the C$_{60}$ film are 24.4 eV/nm and
24.8 eV/nm respectively. As the proton energy loss is uniform over
the whole depth of the film it is reasonable to assume that the
irradiation damage is uniformly produced over the whole depth of
the film. We have used the total thickness of the film in order to
determine the magnetization value in emu/g.

The magnetization curve of Rh-C$_{60}$ \cite{Mak_nat} shows a
tendency to saturate even at an applied field of 2 T, but for the
present irradiated sample a tendency towards saturation can be
seen at high fields. The magnetization at 7 T is about 200 emu/g.
(Similar large  magnetization value (400 emu/g) had been reported
for proton-irradiated HOPG samples \cite{high_mag}). In the
proton-irradiated C$_{60}$ films, although defects are created,
the observation of ordered periodic lattice fringes in the
irradiated sample and corresponding TED pattern indicate that
irradiation did not cause disintegration of C$_{60}$ cage leading
to amorphization.

Regarding the mechanism for the formation of magnetic state in
all-carbon systems, among others, the defect mediated mechanism
appears to be the most general one. The defect-mediated mechanism
has been addressed in a number of publications \cite{Mak_nat,
Wood_jpcm,topology,theory_arxiv} Although the details may be
different for different carbon systems (graphite, polymeric
fullerene, nanotubes etc.), the common feature is the presence of
undercordinated atoms, such as vacancies \cite{dft}, and atoms in
the edges of graphitic nanofragments \cite{esq1,graph1,nanogr1}.
Ion irradiation of any materials generates vacancies. The
enhancement in magnetization observed in H$^+$-irradiated C$_{60}$
samples may be due to defect moments from vacancies and/or
deformation and partial destruction of fullerene cage. The HRTEM
image in Fig. 2 (c) points to this possibility. There are reports
\cite{esq1,graph1,nanogr1} that nanographitic fragments can
trigger ferromagnetism. Formation of graphitic nanofragments in
ion-irradiation of fullerene cannot be ruled out. Further studies
like estimating the presence of nanographitic fragments and carbon
vacancies in such systems will provide deeper insight into the
understanding of origin of magnetism in the ion-irradiated
C$_{60}$ films.

According to a recent density functional study \cite{dft} of
magnetism in proton irradiated graphite \cite{Esqu_prl91}, it is
shown that H-vacancy complex plays a dominant role for the
observed magnetic signal. For a dose of 10 $\mu$C the predicted
signal is 0.8 $\mu$emu which is in agreement with the experimental
signal \cite{Esqu_prl91}. The implanted proton dose in our sample
is 77 $\mu$C ( 6 $\times$ 10$^{15}$ ions cm$^{-2}$) and all the
protons are buried into silicon. So far we have not come across
any report showing magnetic ordering in proton irradiated silicon.
Even if we assume the same kind of magnetism due to H-vacancy
complex in proton-irradiated silicon as in proton-irradiated
graphite, the expected magnetic signal would be three orders of
magnitude smaller than our observed result. Considering the above
fact we can safely ignore the contribution of implanted protons in
the Si substrate to the observed magnetism, which is predominantly
due to atomic displacements caused by energetic protons while
passing through the film.

In conclusion, we have observed ferromagnetic-like behavior in 2
MeV proton-irradiated C$_{60}$ films. Magnetism in this irradiated
films arises due to atomic displacements caused by the energetic
protons as they pass through the film. Possible sources of
magnetization are isolated vacancies, vacancy clusters or
formation of nanographitic fragments. Further investigations are
necessary to estimate their relative contribution.

We thank Prof. S. N. Behera for his important suggestions and a
critical reading of the manuscript. Also we thank S. Rath and
Prof. S. N. Sahu for Raman measurements and T. R. Rautray for PIXE
data acquisition.

\end{document}